\documentclass[epj]{svjour}

\usepackage{latexsym}
\usepackage{graphicx}
\usepackage{amsmath}
\usepackage{epsfig}
\usepackage{subfigure}  
\usepackage{verbatim}   
\usepackage{color}      
\usepackage{hyperref}   %

\begin{document}
\def\bx{{\bf x}}
\def\bu{{\bf u}}
\def\bv{{\bf v}}
\def\bk{{\bf k}}
\def\bp{{\bf p}}
\def\bq{{\bf q}}
\def\br{{\bf r}}
\def\bx{{\bf x}}
\def\bbf{{\bf f}}

\def\eps{\varepsilon}
\def\epsCr{\varepsilon_\mathrm{cr}}
\title{On the Vortex Dynamics in Fractal Fourier Turbulence}
\author{Alessandra S. Lanotte\inst{1}\thanks{\emph{Contact author:}
    a.lanotte@isac.cnr.it} \and Shiva Kumar Malapaka\inst{2,3} \and
  Luca Biferale\inst{3}
}                     

\institute{ISAC-CNR and INFN Sez. Lecce, 73100 Lecce,
  Italy \and IIIT-Bangalore 26/C, Electronics City, Hosur Road, Bangalore -
  560100, India \and Dept. of Physics and INFN, University of Rome Tor Vergata,
  Via della Ricerca Scientifica 1, 00133 Roma, Italy}
\date{Received: date / Revised version: date}

\abstract{Incompressible, homogeneous and isotropic turbulence is
  studied by solving the Navier-Stokes equations on a reduced set of
  Fourier modes, belonging to a fractal set of dimension $D$. By
  tuning the fractal dimension parameter, we study the dynamical
  effects of Fourier decimation on the vortex stretching mechanism and
  on the statistics of the velocity and the velocity gradient
  tensor. In particular, we show that as we move from $D=3$ to $D \sim
  2.8$, the statistics gradually turns into a purely Gaussian one. This
  result suggests that even a mild fractal mode reduction strongly
  depletes the stretching properties of the non-linear term of the
  Navier-Stokes equations and suppresses anomalous fluctuations. \\
}
\authorrunning {Lanotte et al.}
\titlerunning {On the Vortex Dynamics in Fractal Fourier Turbulence}
\maketitle
\section{Introduction}
\label{intro}
A distinctive feature of three-dimensional fully developed turbulent
flows is the presence of bursty fluctuations in the velocity increment
statistics over a wide range of scales, a phenomenon called {\it
  intermittency} \cite{frisch}. The statistical signature of such
fluctuations is the violation of the self-similar Kolmogorov theory in
the inertial range of scales.\\ While Eulerian
\cite{frisch,Arneodo1996,Sreeni1997} and Lagrangian
\cite{Mo2001,PoF2005,Xu2006,PRL2008,rev_toschi_bode} observations
leave no doubt about the existence of intermittency, a theoretical
framework explaining its origin and its relation to the direct cascade
of kinetic energy is still lacking. The question is fundamental
\cite{frisch,Kr71,Kr74} and practical since modeling relies on
assumptions invoking scaling invariance and scale-by-scale energy
budgets~\cite{MK00,SreFa06}. During the formation of strong
fluctuations, large spatial structures create thin vorticity layers or
filaments under both the action of shearing and stretching. Vortex
stretching is essentially a process of interaction of vorticity and
strain and is an important mechanism for understanding both
intermittency as well as energy cascade in a turbulent flow
\cite{frisch,T99}. Its role can be quantified in experiments and
numerical simulations, while closure approximations
\cite{K61,orszag77} as well as phenomenological models \cite{frisch}
for homogeneous and isotropic turbulence fail to account for vortex
structures. \\In this paper, we propose to further investigate the
relation between intermittency and vortex stretching by a novel
approach to three dimensional turbulence. This consists in numerically
solving the Navier-Stokes equations on a multiscale sub-set of Fourier
modes (also dubbed Fourier skeleton), belonging to a fractal set of
dimension $D \le 3$ \cite{frisch2012,LBBMT2015}. For $D=3$, the
original problem is recovered. This implies that the velocity field is
embedded in a three dimensional space, but effectively possesses a
number of Fourier modes that grows slower as $D$ decreases: in
particular, in the Fourier space the number degrees of freedom inside
a sphere of radius $k$ goes as $\#_{dof}(k) \sim k^{D}$.\\ Attempts to
study homogeneous and isotropic $D$-di\-men\-sio\-nal turbulence, with
$2 \le D \le 3$, are not new (see \cite{FF1978}), and were mostly
inspired by statistical mechanics approaches to hydrodynamics: the
idea is to find non-integer dimensions where closures, compatible with
Kolmogrov 1941 theory, can be satisfactorily used. Equilibrium
statistical mechanics in relation to Galerkin-truncated,
three-dimensional Euler equations has been also used to study
three-dimensional turbulence (see pioneering works by Lee \cite{Lee}
and Hopf \cite{Hopf}). In particular, recent numerics
\cite{brachet2005}) of the Euler eq. with a large but finite number of
Fourier modes has interestingly shown that in the relaxation towards
the equilibrium spectrum, large-scale dynamics exhibits a Kolmogorov
spectrum. This suggests that relevant features of the {\it turbulent
  cascade} can be studied in terms of the {\it thermalization
  mechanism} \cite{ray2015}. \\In \cite{frisch2012}, the idea of
Galerkin truncation was adopted to investigate, in two-dimensional
turbulence, the link between the inverse energy cascade and
quasi-equilibrium Gibbs states with Kolmogorov spectrum, when the
dynamics is restricted on a fractal set with $D=4/3$
\cite{lvov}. Finally, the idea of changing the "effective dimension''
between $D=2$ and $D=3$ has been explored within shell models of
turbulence, by modifying the conserved quantities of the system
\cite{GJY2002}.\\ More recently, in \cite{LBBMT2015}, fractally
Fourier decimated Navier-Stokes equations were studied for the first
time in the range $2.5 \le D \le 3$.  Two main results emerged: (i)
average fluctuations are mildly affected by the decimation, since the
kinetic energy spectrum exponent gets a correction linear in the
codimension $3-D$, i.e. $E^{D}(k) \propto k^{-5/3 + (3-D)}$; (ii)
differently, large fluctuations are severely modified, since the
probability density function (PDF) of the vorticity becomes almost
Gaussian already at $D=2.8$. \\ Here, we study more extensively the
velocity increment statistics, and the vortex streching mechanism, as
quantified by the statistics of second and third order invariants of
the velocity gradients tensor. We show that it is significantly
changed as we move from $D=3$ to $D=2.5$, with the evidence of the
intermittent behaviour almost vanishing even for a tiny decimation,
i.e., for $D \simeq 2.98$. This leaves a distictive mark on the
vorticity field: the filamentary structure at $D=3$ is replaced by a
proliferation of small grains of vorticity populating all regions of
the flow (as shown in Figure~\ref{fig:1}).\\ In Section~\ref{sec:2},
  we describe the equations and numerical methods used to generate the
  dataset, as well the statistical approach adopted to analyse it. In
  Section~\ref{sec:3}, we first discuss few results about the
  statistical behaviour of velocity fluctuations and the spectral
  properties of Fourier decimated turbulence. Then, we focus on the
  small-scale statistics by analysing the velocity gradient tensor
  statistics. Finally we provide some conclusions in
  Section~\ref{sec:4}.
\begin{table*}
\begin{center}
\scriptsize{
\begin{tabular}{llllllllllll}
\hline $D$ & $3$ & $3$ & $2.999$ & $2.99$ & $2.99$ & $2.98$ & $2.98$ &
$2.8$ & $2.8$ & $2.5$ & $2.5$\\ $N$ & $1024$ & $1024$ & $1024$ & $1024{\it (*)}$ &
$2048$ & $1024$ & $2048$ & $1024$ & $1024$ & $1024$ & $1024{\it(*)}$ \\ $\nu$ & $6.e-4$
& $8.e-4$ & $6.e-4$ & $6.e-4$ & $2.e-4$ & $6.e-4$ & $2.e-4$& $6.e-4$& $8.e-4$ & $6.e-4$ & $1.5e-4$\\ 
$M_r$ & $100\%$ & $100\%$ & $99\%$ & $93\%$ & $92\%$ & $87\%$ & $85\%$ & $25\%$ & $25\%$ & $3\%$ & $3\%$\\ 
$\eta$ & $0.75$ & $0.95$ & $0.75$ & $0.95$ & $0.70$ & $0.75$ & $0.70$ & $0.40$ &
$0.90$& $0.65$ & $0.2$ \\ ${\cal N}_{T}$ & $10$ & $10$ & $10$ & $11$ & $10$ & $11$ &
$7$ & $10$& $10$ &$10$ & $10$\\ \hline
\end{tabular}}
\caption{Direct Numerical Simulation parameters. The fractal dimension
  $D$; the grid resolution per spatial direction $N$; the viscosity
  $\nu$; the percentage $M_r$ of Fourier modes that survives to the
  decimation action; Kolmogorov length scale $\eta$ in grid spacing
  units, where the grid spacing is $\Delta x= 2\pi/N$; the number of
  steady state large-scale eddy-turnover-times collected for
  statistical analysis, ${\cal N}_{T}$. For all simulations, $L=2\pi$
  is the size of the system. Runs labeled with an asterisk {\it (*)}
  have been performed with two different realisations of the fractal
  mask. Note that the reference wavenumber $k_0$ entering the definition of
  the probability, $h_k= (k/k0)^{D−3}$, is eual to 1.}
\label{table:param}
\end{center}
\end{table*}

\section{Model Equations and Methods}
\label{sec:2}
\subsection{The Navier-Stokes equations on a Fractal Fourier set}
\label{sec:2.1}
Let us define $\bv(\bx,t)$ and $\hat{\bu}(\bk,t)$ as the real and Fourier
space representation of the velocity field in $D=3$, respectively. We
then introduce a decimation operator ${\cal P}^{D}$ that acts on the
velocity field as:
\begin{equation}
\label{eq:decimOper}
{\bf v}^{D}(\bx,t)= {\cal P}^{D}{\bf v}(\bx,t)=\hspace{-1mm}
\sum_{{\bf k}\in {\cal Z}^3}\hspace{-1mm} e^{i {\bf k \cdot x}}\,\gamma_{\bf k}\hat{\bu}(\bk,t)\,.
\end{equation}
 Here $\bv^{D}(\bx,t)$ is the decimated velocity field.  \\ In this equation
 $\gamma_{\bk}$ represent random numbers that are quenched in time and
 are determined as :
\begin{equation}
\label{eq:theta}
\gamma_{\bf k} =
\begin{cases}
1, & \text{with probability}\ h_k\,, \\
0, & \text{with probability}\ 1-h_k, k\equiv|{\bf k}|\,.
\end{cases}
\end{equation}
The choice for the probability $h_k \propto (k/k_0)^{D-3}$, with $0< D
\le 3$ and $k_0$ a reference wavenumber, ensures that the dynamics is
isotropically decimated to a $D$-dimensional Fourier set. The factors
$h_k$ are chosen independently and preserve Hermitian symmetry
$\gamma_k= \gamma_{-k}$ so that ${\cal P}^{D}$ is self-adjoint as was
described in \cite{frisch2012}. The Navier-Stokes equations for the
decimated velocity field $\bv^{D}(\bx,t)$ are then defined as:
\begin{equation}
\label{eq:decimNS}
\partial_t {\bf v}^{D} = {\cal P}^{D}N({\bf v}^{D},{\bf v}^{D}) + 
  \nu \,\nabla^2 \bv^{D} + {\bf F}^{D}\,. 
\end{equation}
Here $N({\bf v},{\bf v})= -{\bf v}\cdot {\bf {\nabla v}} + {\bf
  \nabla}p$ is the non-linear term of the NS equation. Equation
~\ref{eq:decimNS} conserves both energy and helicity in the inviscid
and unforced limit, exactly as in the original (non decimated) problem
with $D=3$; ${\bf F}^{D}$ is the large-scale forcing, injecting kinetic
energy in the system. The notation above, ${\cal P}^{D}N({\bf
  v}^{D},{\bf v}^{D})$, is to imply the fact that the non-linear term
is projected, at every time iteration, on the quenched fractal set, so
that its dynamical evolution remains on the same Fourier skeleton at
all times. Similarly, the initial condition and the external forcing
are defined on the same fractal set of Fourier modes. \\In the sequel,
for the sake of simplicity, we shortly refer to $\bv(\bx)$ and
$\hat{\bv}(\bk)$ as the real and Fourier space representation of the
solution of the decimated Navier-Stokes equations (\ref{eq:decimNS}).
\begin{figure}
\subfigure{\includegraphics[width=0.225\textwidth]{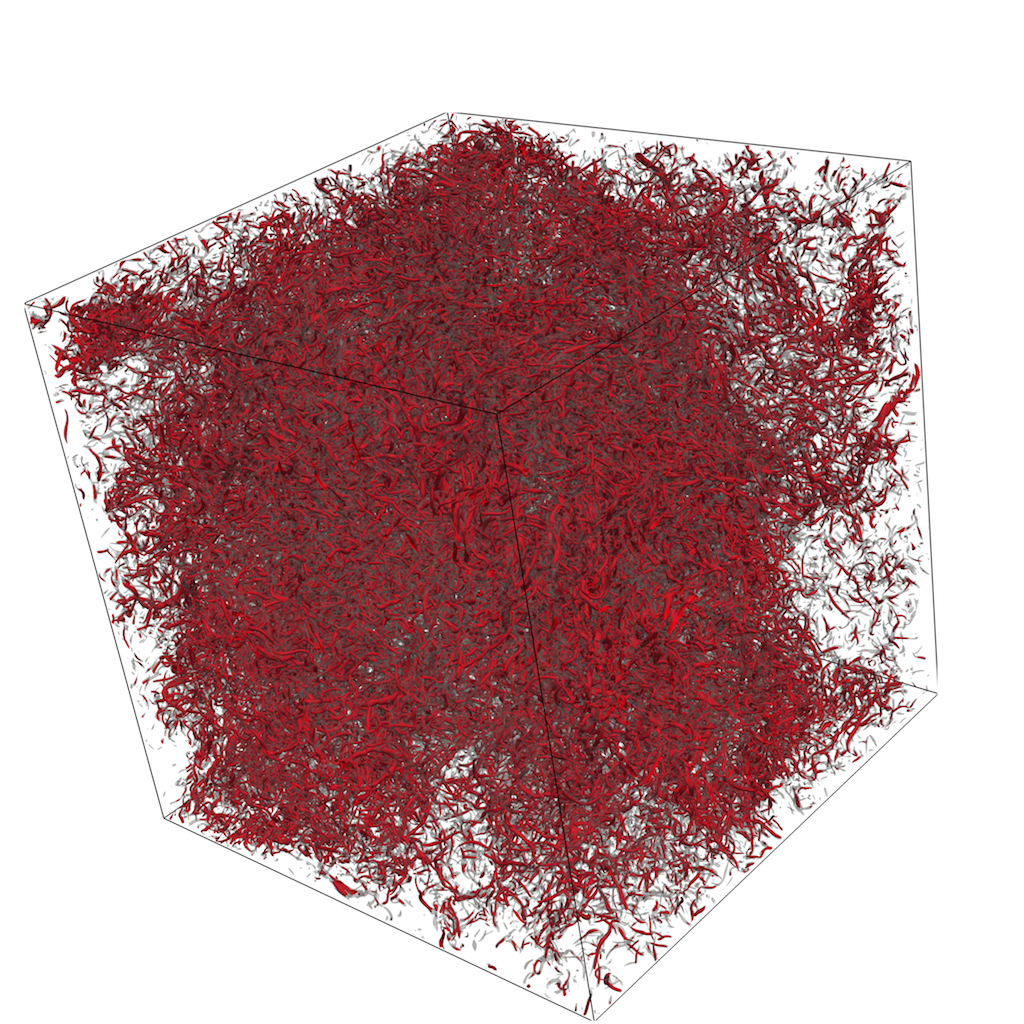}}
\subfigure{\includegraphics[width=0.225\textwidth]{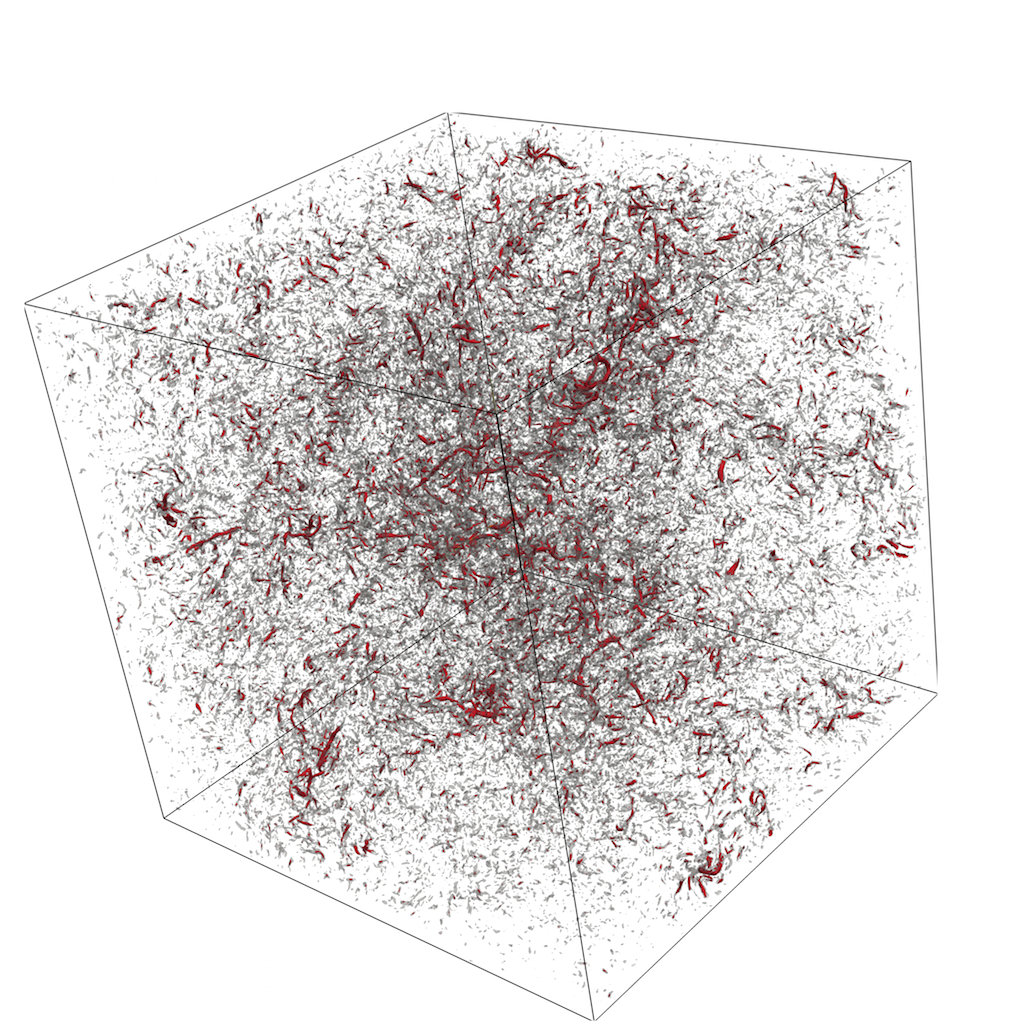}}\\
\vspace{0.3cm}
\hspace{0.2cm}
\subfigure{\includegraphics[width=0.2\textwidth]{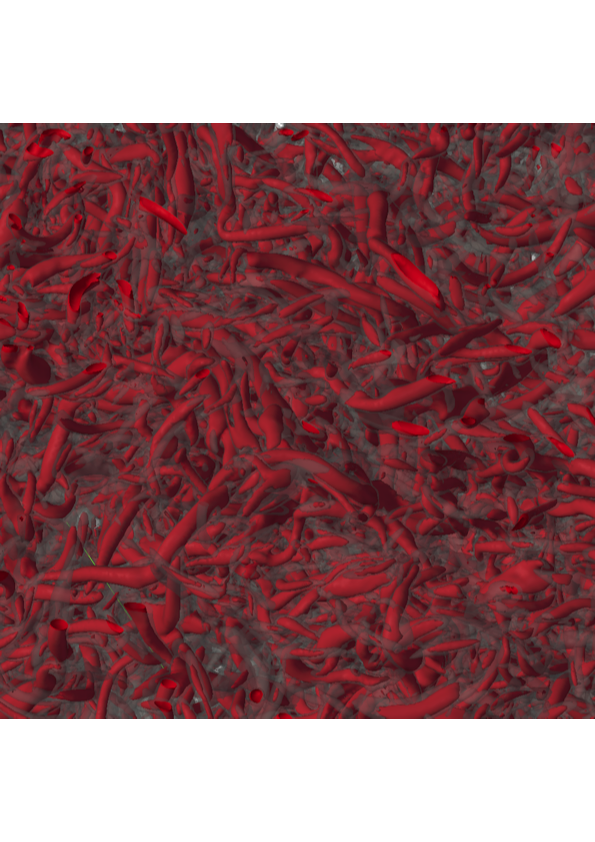}}
\hspace{0.2cm}
\subfigure{\includegraphics[width=0.2\textwidth]{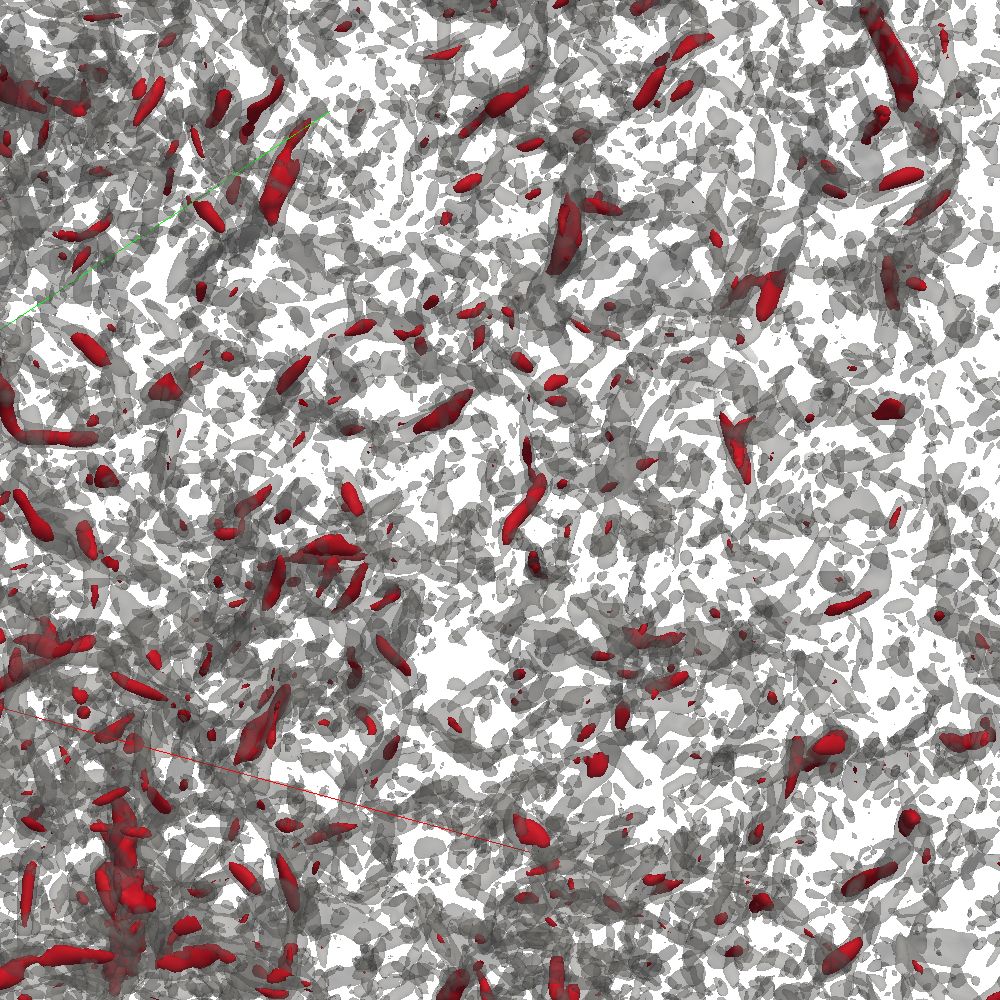}}
\caption{Plot of the most intense vortical structures. (top) A
  snapshot of the turbulent flow with $D=3$ (left) and with $D=2.98$
  (right). Isosurfaces of the $Q$ invariant (see text) of the velocity
  gradient tensor are plotted: values $Q/Q_{rms}=1$ (grey) and
  $Q/Q_{rms}=2$ (red). (bottom) A zoom in the top snapshots highlights
  details of the small scales. Data refer to runs with $N=1024$.}
\label{fig:1}
\end{figure}
\begin{figure}
\subfigure{\includegraphics[width=0.225\textwidth]{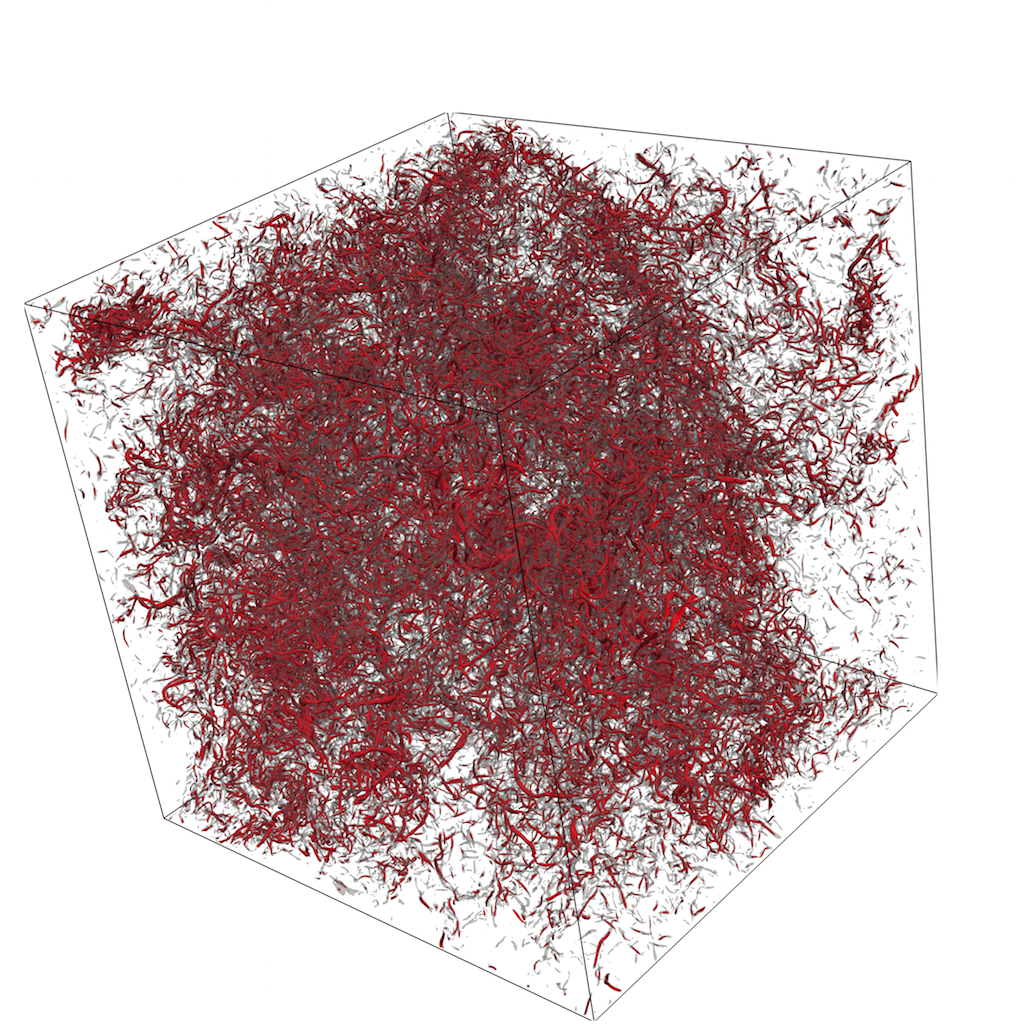}}
\hspace{0.2cm}
\subfigure{\includegraphics[width=0.2\textwidth]{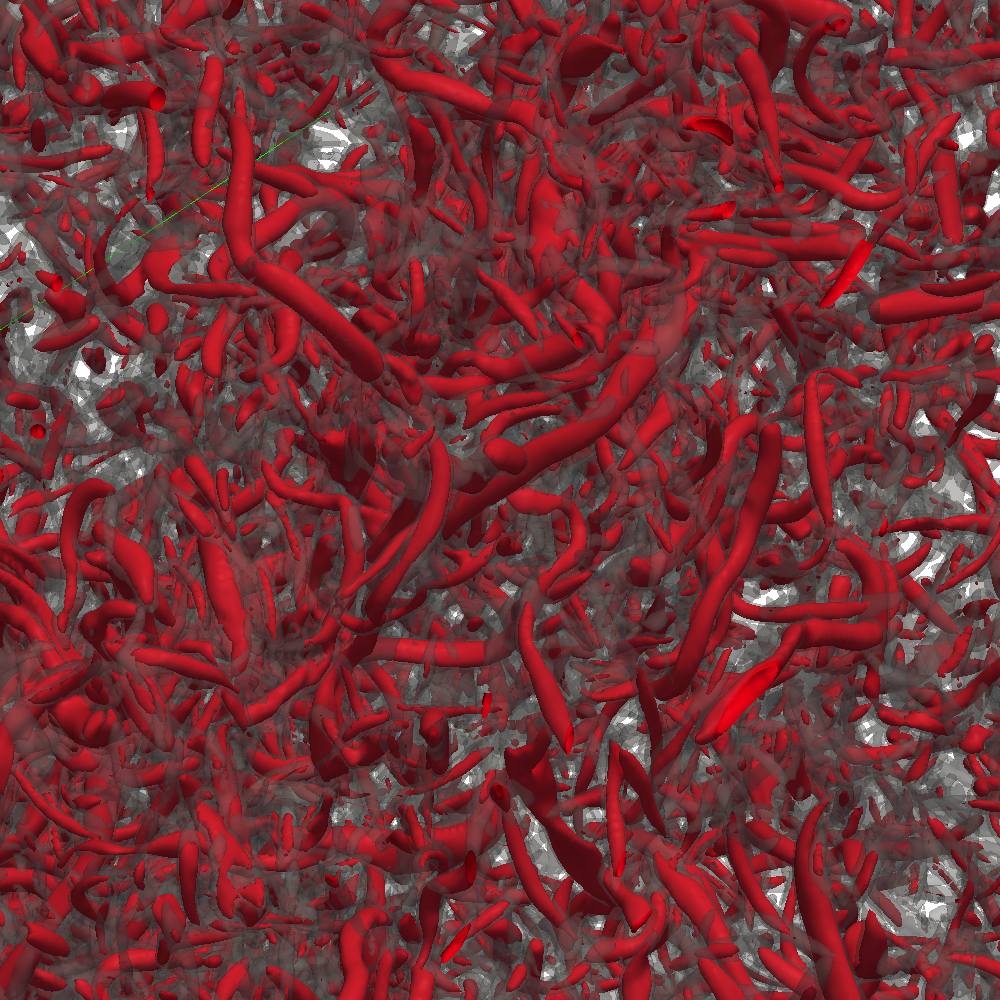}}
\caption{Isosurfaces of the $Q$ invariant as above. Here we show what
  is left of the turbulent flow with $D=3$ ($N=1024$), after applying
  the static mask with dimension $2.98$. The full volume on the left;
  a zoom highlighting small-scale features, on the right.}
\label{fig:1bis}
\end{figure}

\subsection{Direct Numerical Simulations Set-up}
\label{sec:2.2}
 We solved equations~(\ref{eq:decimNS}) on a regular, periodic volume
 with $N=1024^3$ and $2048^3$ grid points, by adopting a standard
 pseudo-spectral approach fully dealiased with the two-thirds rule;
 time stepping is done with a second-order Adams-Bashforth scheme. A
 large-scale forcing ${\bf F}$ keeps the total kinetic energy constant
 in a range of shells, $ 0.7 \le |{\bf k}| < 1.7$, leading to a
 statistically stationary, homogeneous and isotropic flow
 \cite{pope}. For each run, we generated a {\it mask}, that is kept
 quenched throughout the numerical simulation. We performed several
 runs at changing the fractal dimension $2.5 \le D \le 3$, the spatial
 resolution, the viscosity and also the realization of the fractal
 quenched mask. The case for $D=3$ is also referred as standard
 case. We summarise in Table\ref{table:param} the relevant parameters
 of the numerical experiments performed. \\ We mention that an {\it a
   posteriori} projection on a set of fractal dimension $D$ can also
 be obtained by applying, in Fourier space, the mask on snapshots of
 the velocity field which is solution of the original
 three-dimensional Navier-Stokes equations. This is a {\it static}
 fractal Fourier decimation, whose effect can be compared to that of
 the {\it dynamical} decimation, in the statistical analysis.

\section{Results and Discussion}
\label{sec:3}
We start our analysis by considering a visualisation of the most
intense vortical structures, revealing the effect of decimation on
turbulent flows. In Figure~\ref{fig:1}, we plot isosurfaces of the $Q =
Tr[{\bf A}^2]$ invariant of the velocity gradient tensor, $A_{ij} =
\partial_i v_j$. The Q-criterion is based on the observation that
$$
Q= \frac{1}{2}(\Omega_{ij}\Omega_{ij} - S_{ij}S_{ij});  
$$ where the vorticity tensor is $\Omega_{ij} = 1/2 (\partial_i v_j
-\partial_j v_i)$ and the rate-of-strain tensor is $S_{ij} = 1/2
(\partial_i v_j +\partial_j v_i)$. Therefore, flow regions where $Q$
is positive identify positions where the strength of rotation
overcomes strain. These are the best candidates to be considered as
vortex iso-surfaces \cite{Dubief}. From Figure~\ref{fig:1}, we see
that the $D = 3$ case shows a large number of structures of both large
and small-scale vortex filaments. The decimated case with $D = 2.98$
clearly differs because structures are smaller and less elongated,
also they are much less abundant, indicating a less intermittent
spatial distribution of structures. We stress that fractal decimation
has non-trivial dynamical effect, which differs from the simple action
of a vector projection in Fourier space. To make this immediately
clear, we plot in Figure~\ref{fig:1bis} the $Q$ isosurfaces obtained
after applying the {\it a posteriori}, {\it static} mask of dimension
$D=2.98$ on the $D=3$ turbulent velocity field. While the {\it static}
decimation {\it simply} removes velocity fluctuations at specific
wavenumbers, the dynamical action of the Fourier decimation provokes a
complete reorganization of the flow structures.
\begin{figure}
\begin{center}
\includegraphics[width=8cm]{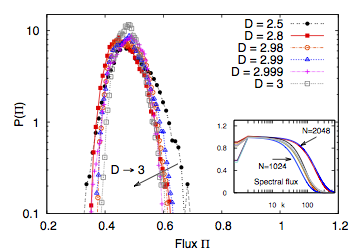}
\caption{(Main body) The lin-log plot of the probability density
  function of the spectral flux $P(\Pi(k))$, normalised to unitary area,
  for runs with different $D$; resolution is $N=1024$. (Inset) the
  comparison of the spectral fluxes obtained at the two resolutions,
  $N=1024$ and $N=2048$, and for different fractal dimensions $D$.}
\label{fig:flux}       
\end{center}
\end{figure}

\subsection{Velocity field statistics}
\label{sec:3.1}
We consider the statistics of mean turbulent fluctuations by analysing
the spectral behaviour of the kinetic energy spectrum and energy flux,
when the fractal dimension $D$ is varied. 
\begin{figure}
\begin{center}
\includegraphics[width=8cm]{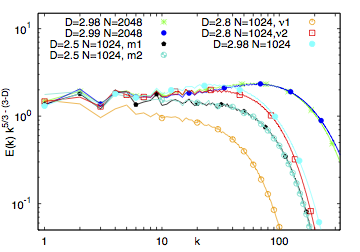}
\caption{Log-log plot of the compensated kinetic energy spectra, $E(k)\,
  k^{5/3 - (3 -D)}$, obtained at the two resolutions, $N=1024$ and
  $N=2048$, and for different fractal dimensions $D$. We only report
  those cases for which the linear correction to the exponent, due to
  the fractal decimation, is appreciable. At resolution $N=1024$, for
  $D=2.5$ label $m1$ and $m2$ indicate two sets of numerical
  simulations with different masks. At resolution $N=1024$, for
  $D=2.8$, label $\nu_1$ and $\nu_2$ indicate runs with different
  viscosities equal to $6.e-4$ and $8.e-4$, respectively: for these
  runs the masks are also different.}
\label{fig:spectra}   
\end{center}
\end{figure}
From the DNS data, the energy spectrum is measured by angular averages
on Fourier-space unitary shells,
\begin{equation}
E^{D}(k) \equiv \sum_{k\le|\bk|\le k+1} \langle v_i(\bk) v_i(\bk)^*\rangle\,,
\label{eq:spectrum_discrete}
\end{equation}
where the asterisk is for complex conjugation. The energy flux
$\Pi^{D}(k)$ through wavenumber $k$ due to the nonlinear transfer is
measured as
\begin{equation}
\Pi^{D}(k) \equiv \sum_{|\bq|< k} \sum_{\bp}\langle \bv^*(-\bq -\bp) \cdot \left[\bq \cdot (\bv(\bq) \bv(\bp))\right]\rangle\,.
\label{eq:flux_discrete}
\end{equation}
As reported in \cite{LBBMT2015}, a dimensional argument can be built
up to quantify possible modification of the exponent of the kinetic
energy spectrum due to fractal decimation. It relies on two
assumptions: (i) scaling invariance of the velocity fluctuations in
the inertial range of scales; (ii) the existence of a constant
(k-independent) spectral energy flux in the inertial range. We give
the former for granted, since intermittency manifests only in terms of
a tiny anomalous correction in the energy spectrum of
three-dimensional turbulence \cite{IsGoKa09}. 
\begin{figure}
\begin{center}
\includegraphics[width=8cm]{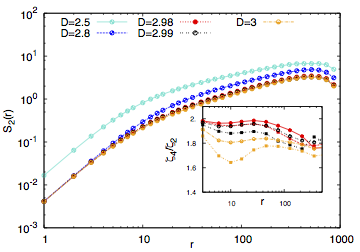}
\caption{(Main body) Log-log plot of the longitudinal second order
  structure function for different fractal dimensions $D$. Data are at
  resolution $N=1024$. (Inset) Log-lin plot of the local slopes in ESS
  of the 4th-order longitudinal (filled circles) and transverse
  (filled boxes) structure functions in terms of the second order one,
  versus the separation scale $r$. Color symbols are the same of the
  main body: lower curves are for $D=3$, middle curves for $D=2.99$
  and upper curves for $D=2.98$.}
\label{fig:S2}    
\end{center}
\end{figure}
As for the latter, we plot in the inset Figure \ref{fig:flux}, the
kinetic energy flux $\Pi(k)$ through wavenumber $k$ for DNS with
different fractal decimation $D$, and for both resolutions. It is
clear that even in the presence of a strong reduction of degrees of
freedom, an average constant flux in Fourier space is observed and a
cascade of kinetic energy takes place.\\ We also quantify the
fluctuations in the Fourier space energy transfer, by plotting in
Figure~\ref{fig:flux} the probability density function of the spectral
flux. The PDF is calculated measuring the fluctuations of $\Pi(k)$ for
wave-numbers in a limited range $k\in[3:30]$, corresponding to the
constant transfer region where the flux has a plateau. We can observe
that the fluctuations of the spectral transfer tend to be of the same
amplitude, when $D$ is changed. Only for the case of a very strong
decimation with $D=2.5$, we notice the presence of slightly larger
fluctuations due to the fact the number of triads in the dynamical
process are less and less as $D$ increases.\\
\begin{figure}
\begin{center}
\includegraphics[width=8cm]{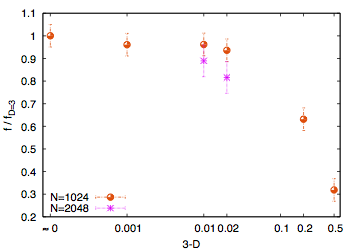}
\caption{Log-lin plot of the normalised drag reduction coefficient
  $f/f_{D=3}$ versus the dimension deficit $3-D$. Error bars are
  estimated from statistical fluctuations. }
\label{fig:friction}      
\end{center}
\end{figure}
Figure \ref{fig:spectra} shows the compensated kinetic energy spectra,
$E(k) k^{5/3-(3-D)}$: fractal decimation acts to make the spectrum
shallower, and the correction is small, being linear in the dimension
deficit $3-D$ \cite{LBBMT2015}. Such correction is clearly negligible
for those runs at $D \in [2.98,3)$, but starts to be appreciable at
  $D=2.8$. At $D=7/3$ the spectrum would in principle become divergent
  at small wavenumbers. However at such fractal dimension, about less
  than $1\%$ of the modes would survive for the present resolutions,
  almost annihilating the role of the non-linear transfer and making
  the energy dissipation almost in a direct balance with the energy
  injection. We note that, recently, the Burgers
    equation decimated on a fractal Fourier set of dimension $D \le 1$
    was numerically studied \cite{BBFR2016}. Results obtained at fixed
    $D$ and for larger and larger values of the Reynolds number,
    suggest that Fourier decimation is a singular perturbation for the
    spectral scaling properties. Should this happen for the
    Navier-Stokes equations also, it is something to explore.\\It is
    also important to notice that fractal Fourier decimation has a
    strong effect on the vortex stretching mechanism: this implies
    that the energy bottlenecks \cite{F1994,Lohse1995,Frisch2008},
    generally observed at high Reynolds numbers in $D=3$ turbulence,
    might become less and less important for $D<3$.\\

  We now consider the Fourier transform of the energy spectrum, giving
  the second order velocity longitudinal structure function
  $S^{(2)}_L\equiv \langle [(\bv(\bx + \br) - \bv(x))\cdot
    \hat{\br}]^2 \rangle$. This is plotted in Figure~\ref{fig:S2}. By
  decreasing the fractal dimension $D$, the curves become less and
  less steep approaching the Kolmogorov dimensional scaling $S^{(2)}_L
  \simeq (r/L)^{\zeta_2}$, linearly corrected by the dimension deficit
  $\zeta_2=2/3-(3-D)$. Note that the run at $D=2.5$ has a smaller
  kinematic viscosity, hence a larger kinetic energy.\\
\begin{figure*}
\begin{center}
\includegraphics[width=14cm]{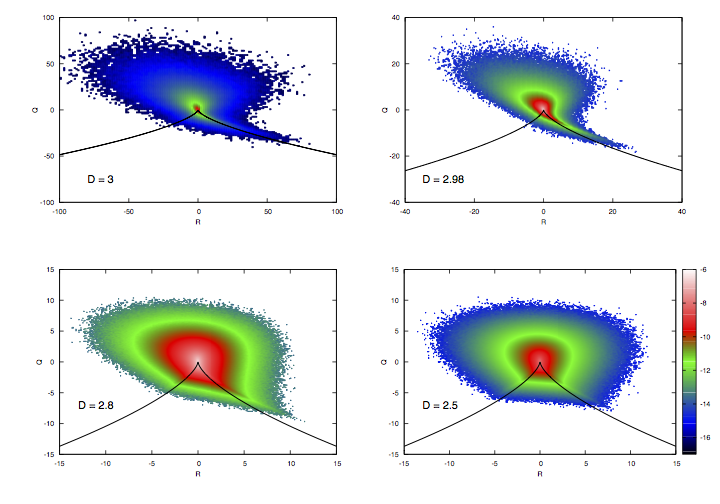}
\caption{Isosurfaces of the joint probability density function
  $P(Q^*,R^*)$, where $Q^*=Q/\langle Q^2 \rangle^{1/2}$ and
  $R^*=R/\langle Q^2 \rangle^{3/2}$. Isosurfaces are equispaced
  logarithmically. Note that when the fractal dimension $D$ is
  lowered, the range of fluctuations for the $Q-R$ variables is
  substantially modified, and become much smaller. The black line is
  the Viellefosse line $L\equiv 27/4 R^2 - Q^3 = 0$.}
\label{fig:2}     
\end{center}
\end{figure*}
It is interesting to check how longitudinal and trasversal structure
functions approach the dimensional scaling. Indeed it has been
observed that, in statistically homogeneous and isotropic turbulence,
longitudinal and trasversal structure functions do have different
anomalous corrections \cite{PhysD2008}, contrary to what would be
expected on the basis of symmetry arguments
\cite{BifProc2005}. Whether this is an effect due to finite-Reynolds
numbers or not remains an open issue. In the inset of
Figure~\ref{fig:S2}, we plot the local slopes of the longitudinal and
transverse fourth order structure functions, by using the so-called
extended self-similarity technique (ESS)\cite{ess} (see also
\cite{ray2010} for a recent discussion of the topic). To be more
precise, we consider the ratio of the scaling exponent of the fourth
order longitudinal (transverse) structure function to that of the
second order longitudinal (transverse) one:
\begin{equation}
\frac{\zeta_4}{\zeta_2}(r)= \frac{d\log S^{(4)}_{L,T}(r)}{d\log S^{(2)}_{L,T}(r)}\,,
\label{eq:ess}
\end{equation}
where the longitudinal structure functions are $S^{(n)}_L(r) \equiv
\langle [\delta_r {\bv} \cdot \hat {\bf r}]^n \rangle$ with $\delta_r
        {\bv} = {\bv}({\br}) - {\bv}(0)$, and the purely transverse
        structure functions are $S^{(n)}_T(r) \equiv \langle [\delta_r
          v(\hat{\br}_T)]^n \rangle$ and $\hat{\br}_T \cdot \bv
        =0$. For $D=3$, the curve for the transverse structure
        function is well below that of the longitudinal one, hence the
        anomalous correction is larger for the former than for the
        latter \cite{Gotoh2002,Benzi2010}. When the fractal dimension
        is decreased, $D<3$, such difference also diminishes but the
        two curves remain separated even when approaching
        Gaussianity. This might suggest that such difference in the
        standard $D=3$ turbulence is not due to finite Reynolds
        effects but it is genuine, and different anomalous corrections
        characterize longitudinal and transverse structure
        functions.\\

A different way to quantify the effects of fractal decimation, beyond
scaling properties, is to look at the ratio of the kinetic energy
input to the resulting kinetic energy throughput in the turbulent
flow. An adimensional measure can be defined as \cite{BCM2005}
\begin{equation}
f_{d}\equiv \frac{F\,L}{\langle v \rangle^2} = \frac{2 \epsilon L}{v^3_{rms}} \,
\end{equation}
where $F$ quantifies the amplitude of the external forcing and $L$ is
the large-scale of the flow; moreover from the definition of the
energy input $\epsilon \equiv 1/2 \langle {\bf F} \cdot \bv \rangle$
we have the dimensional relation $F\simeq 2 \epsilon /\langle
v^2\rangle^{1/2}$. In this terms, $f_{d}$ is a drag coefficient which
can be measured when the fractal dimension is varied. From the results
plotted in Fi\-gu\-re~\ref{fig:friction}, it appears that fractal
decimation reduces the drag in the flow with respect to the standard
case with $D=3$. Moreover there is a Reynolds dependence: we observe a
larger reduction for larger Reynolds numbers. This sort of drag
reduction is accompanied, as we will see below, by a restructuring of
the flow, since regions characterised by intense vortical stretching
almost disappear as $D$ is decreased.

\subsection{Velocity gradient tensor statistics}
\label{sec:3.2}
As discussed before, the vortex stretching mechanism can be quantified
by measuring the statistical behaviour of the invariants of velocity
gradient tensor ${\cal{\bf A}}_{ij}= \partial v_i/\partial x_j$ (for a
detailed discussion, see \cite{CPC,BMC}). The characteristic equation
$det ({\cal{\bf A}} - \lambda {\cal {\bf I}})=0$ can be written as:
\begin{equation}
\lambda^3 + P \lambda^2 +  Q \lambda + R=0\,.
\label{eq:char}
\end{equation}
For an incompressible flow, of the three tensor invariants only two
are different from zero: $Q = - 1/2 tr[{\cal {\bf A}}^2]$ and $R= -
1/3 tr [{\cal {\bf A}}^3]$.\\ From previous experimental
\cite{Tao2002,Luthi2005} and numerical \cite{Martin1998} studies in
three-dimensional homogeneous and isotropic turbulence, some general
geometric features of the tensor have been highlighted. These are: the
vorticity vector is preferentially ali\-gned with the eigenvector
associated to the intermediate eigenvalue of the strain-rate tensor
${\cal {\bf S}}$; there are two positive and one negative eigenvalues
of the rate-of-strain, ${\cal {\bf S}}$, such that the associated
local flow structure is an axisymmetric extension; the joint
probability distribution of the two invariants, $P(Q,R)$, has a
typical teardrop shape extending around the so-called
zero-discriminant or Vieillefosse line $L \equiv 27/4 R^2 + Q^3 = 0$
\cite{viellefosse}. The Viellefosse line divides the $Q-R$ plane in
two different regions depending whether the velocity gradient tensor
has three real eigenvalues with $L<0$, or two complex-conjugate and
one real eigenvalues, with $L>0$. This means that $L> 0$ is the region
where vorticity is dominant; the region $L<0$ is strain-dominated
region. The upper left region (with $L> 0$, and one positive real
eingenvalue and two complex-conjugates ones) is the statistical
signature of the vortex stretching dominating the turbulent flow,
while long right tail in the lower right side ($L<0$ and two positive
and one negative real eingenvalues) is associated with intense
elongational strain.\\In the original $D=3$ problem, the temporal
evolution of the velocity gradient tensor is also of particular
interest beyond the geometrical properties. Indeed a set of exact
equations \cite{viellefosse,cantwell1992}, although not closed, can be derived
ta\-king the gradient of Navier-Stokes equations. By doing so, the
equations for the Lagrangian evolution of the gradient tensor
components is\,
\begin{equation}
\label{eq:gradevol}
d{\cal A}_{ij}/dt = - {\cal A}_{ik}{\cal A}_{kj} - \partial_{ij}p +
\nu \partial^2 {\cal A}_{ij}\,,
\end{equation} 
where $p$ is the pressure divided by the fluid density. The need of a
closure comes from the fact that both the pressure hessian and the
viscous terms are not simply known in terms of ${\bf A}$. Such a set
of equations, which gives insight on the small-scale intermittency,
has been largely investigated, and different model closures have been
proposed
\cite{viellefosse,Girimaji1990,CPS1999,Chevillard2006,Chevillard2008}. \\In
the case of fractal Fourier Navier-Stokes equations, the situation is
however different. Because of the presence of the decimation projector
in the non-linear term, the structure of the equation of motion in
terms of the material derivative is broken. Hence, any closure model
based on the lagrangian evolution of the velocity gradient tensor is
ruled out.\\ In Figure\ref{fig:2}, we plot the joint distribution of
the velocity gradient tensor invariants, $P(Q,R)$ for data at
$N=1024$, since the dataset is richer. As the fractal dimension $D$
decreases, the joint probability looses its asymmetric shape, and
become more and more centered.  Moreover extreme fluctuations becomes
less and less important: tails are reduced in particular for what
concerns the vortex stretching mechanism ($R<0$ and $Q>0$, with $L>0$)
and the region around the Viellefosse tail $R =
(-\frac{4}{27}Q^3)^{1/2}$. On the other hand, small fluctuations for
$Q$ and $R$ become more and more probable. At $D=2.8$, the PDF is
close to the one of Gaussian variables \cite{VanderBos2002,CPS1999},
and the vortex stretching region is strongly depleted.

\section{Conclusions}
\label{sec:4}
In this paper, we have further investigated the effect of random (but
quenched in time) fractal Fourier decimation of the Navier-Stokes
equations in the turbulent regime of direct cascade of energy. This is
a recently introduced technique that allows to study modifications of
the non-linear transfer and vortex stretching mechanisms, by varying
the number of degrees of freedom with a single tuning parameter, i.e.,
the fractal set dimension $D$. Here we have focused on the range $D
\in[2.5:3]$. For specific values of $D$, we have also explored the
dependency on the Reynolds number and on the realization of the
fractal mask. Results here presented do depend on the former, while
they are insensitive to tha letter, within statistical accuracy. Note
that at fixed fractal dimension $D$, in the limit of large Reynolds
numbers, even a small dimension deficit $(3-D)<<1$ would result in an
effective strong decimation at small scales, being $h_k \propto
(k/k_0)^{D-3}$ the probability associated to wavenumbers $k$. This
suggests that the effect of fractal decimation is singular in the
limit $D\rightarrow 3$.\\ Decimation does not alter the energy
cascade of three-dimensional turbulence, meaning that the kinetic
energy flux in Fourier space is independent of the wave-numbers in the
inertial range of scales. Also, fluctuations of the energy flux stay
almost unchanged except for $D=2.5$ where we observe a (mild) increase
in the width of the PDF tails. We interpret this as a manifestation of
the fact that at in such case a very small number of triads is able to
drain the (same) large-scale energy towards small scales: the transfer
hence becomes more {\it difficult} and bursty. \\The second order
moment of velocity increment statistics is weakly affected by fractal
decimation, the correction in the kinetic energy spectrum exponent
being linear in $3-D$. However small-scale statistics is drastically
modified. By studying the velocity gradient tensor statistics, we have
shown that the vortex stretching mechanism is very sentitive to
fractal decimation: it is strogly depleted already for $D$ very close
to three. At $D=2.8$, the statistical signature of vortex stretching
and intense elongational strain disappear from the joint distribution
$P(Q,R)$, which become Gaussian. This implies that
  high order structure functions of the velocity field scale
  dimensionally with the structure function of order two, whose
  exponent is possibly modified by the dimension deficit. Let us
  notice however that the statistics cannot follow an exact
  self-similar behaviour, because some correlation functions are
  anchored to the kinetic energy flux, in particular the third order
  longitudinal structure function must scale linearly.\\From the
results here discussed, many questions arise. Modifications in the
triad-to-triad nonlinear energy transfer mechanism are to be further
investigated. \\ As pointed out in \cite{Waleffe1992}, the energy
transfer mechanism of individual triads might strongly depend on the
helical contents of each interacting mode and on the triad shape, if
this is local or non-local. Fractal Fourier decimation introduces a
systematic change in the relative ratio of local to non-local triads,
tending to deplete the presence of local Fourier interactions for high
wavenumbers. Understanding if the restoration of a non-anomalous
scaling, due to the disappearence of intermittency, and the depletion
of the vortex stretching mechanism are due to this effect needs
further analysis. Finally, let us comment on the fact that fractally
decimated Navier-Stokes equations might be an interesting playground
for more theoretical studies on finite-time singularities. Indeed, by
reducing the number of degrees of freedom when lowering $D$, we
observe that the dynamics tends to be less singular, since anoumalous
fluctuations and events along the Viellefosse tail disappear: this
suggests that solutions of the fractally decimated Navier-Stokes
equations are more regular and hence a good candidate to assess the
presence or not of a blow up at large Reynolds numbers
\cite{constantin,gallavotti}.

\section*{Acknowledgements} 
We acknowledge useful discussions with Roberto Benzi, Uriel Frisch,
Detlef Lohse, Samriddhi Sankar Ray, and Federico Toschi. SKM wishes to
acknowledge COST-Action MP1305 for supporting him to participate in
FLOMAT2015. DNS were done at CINECA (Italy), within the EU-PRACE
Project Pra04, N.806. The research leading to these results has
received funding from the European Union's Seventh Framework Programme
(FP7/2007-2013) under grant agreement No. 339032. We thank
F. Bonaccorso and G. Amati for technical support. We thank the
COST-Action MP1305 for support.

\end{document}